\documentclass[twocolumn,showpacs,preprintnumbers,amsmath,amssymb]{revtex4}

\usepackage{graphicx}
\usepackage{dcolumn}
\usepackage{bm}

\newcommand{\bea}{\begin{eqnarray}}
\newcommand{\eea}{\end{eqnarray}}
\newcommand{\simgt}{\hbox{ \raise3pt\hbox to 0pt{$>$}\raise-3pt\hbox{$\sim$} }}
\newcommand{\simlt}{\hbox{ \raise3pt\hbox to 0pt{$<$}\raise-3pt\hbox{$\sim$} }}

\begin{document}

\preprint{TU--832}

\title{Family Gauge Symmetry and Koide's Mass Formula}

\author{Yukinari Sumino}
\affiliation{Department of Physics, Tohoku University,
Sendai, 980--8578 Japan}%

\date{\today}

\begin{abstract}
Koide's mass formula is an empirical
relation among the charged lepton masses which
holds with a striking precision.
We propose a 
mechanism for cancelling the QED
correction to Koide's formula.
This is discussed in an effective theory with $U(3)$
family gauge symmetry and a scenario in which 
this symmetry is unified with $SU(2)_L$ symmetry
at $10^2$--$10^3$~TeV scale.
\end{abstract}

\pacs{11.30.Hv,12.10.Kt,12.15.Ff,13.20.Eb}
\maketitle

Among various properties of elementary particles,
the spectra of the quarks and leptons exhibit unique patterns, and
their origin still remains as a profound mystery.
Koide's mass formula is an empirical 
relation among the charged lepton masses given by \cite{Koide:1982wm}
\bea
\frac{\sqrt{m_e}+\sqrt{m_\mu}+\sqrt{m_\tau}}{\sqrt{m_e+m_\mu+m_\tau}}
=\sqrt{\frac{3}{2}}\, ,
\label{KoideMF}
\eea
which holds with a striking precision.
In fact, substituting the present experimental values of the charged lepton
masses \cite{Amsler:2008zz}, 
the formula is valid within the present experimental accuracies.
The relative experimental
error of the left-hand side  of eq.~(\ref{KoideMF}) 
is of order $10^{-5}$.

Given the remarkable accuracy with which eq.~(\ref{KoideMF})  holds,
there have been many speculations as to existence of some
physical origin behind
this mass formula \cite{Koide:1982wm,Koide:1983qe}.
Despite the attempts to find its origin, 
so far no realistic model or mechanism has been found
which predicts Koide's mass formula within the required accuracy.
The most serious problem one
faces in finding a realistic model or mechanism 
is caused by the QED radiative correction.
Even if one postulates some mechanism at a high
energy scale 
that leads to this mass relation, 
the charged lepton
masses receive the 1-loop
QED radiative correction given by
\bea
m^{\rm pole}_i = \left[
1+\frac{\alpha}{\pi}\left\{
\frac{3}{4}\log\left(
\frac{\mu^2}{\bar{m}_i(\mu)^2}
\right) +1
\right\}
\right]\,
\bar{m}_i(\mu) \, ,
\label{QED1Lcorr}
\eea
where $\bar{m}(\mu)$ and $m^{\rm pole}$ denote the 
running mass defined in the modified--minimal--subtraction scheme
($\overline{\rm MS}$ scheme) and the pole mass, respectively;
$\mu$ represents the
renormalization scale.
It is the pole mass  that is measured in experiments.
Suppose  $\bar{m}_i(\mu)$ 
(or the corresponding Yukawa couplings $\bar{y}_i(\mu)$)
satisfy the relation
(\ref{KoideMF}) at scale $\mu\gg M_W$.
Then $m_i^{\rm pole}$
do not satisfy the same relation \cite{Li:2006et}: 
Eq.~(\ref{KoideMF})
is corrected by approximately 0.1\%, which is 120 times 
larger than the present experimental error.
Note that this correction
originates only from the term $-3\alpha/(4\pi) \times\bar{m}_i \, \log(\bar{m}_i^2)$
of eq.~(\ref{QED1Lcorr}), since the other terms, which are of the form
${\rm const.}\times\bar{m}_i$, do not affect the relation
(\ref{KoideMF}).
We also note 
that 
$\log(\bar{m}_i^2)$ results from the fact that $\bar{m}_i$
plays a role of an infrared  cut--off in the
loop integral.

The 1--loop weak correction is of the form
${\rm const.}\times\bar{m}_i$ in the leading order of
$\bar{m}_i^2/M_W^2$ expansion;
the leading non--trivial correction is
${\cal O}(G_F \bar{m}_i^3/\pi)$ whose effect can be safely neglected.
Other known radiative corrections are also  negligible.

Thus, 
if there is indeed a physical origin to Koide's mass formula at
a high energy scale,
we need to account for a correction to the relation
(\ref{KoideMF})
that cancels
the QED correction.
Since such a correction is absent up to the scale of
${\cal O}(M_W)$ to our present knowledge, it must originate from a higher scale.
Then, there is a difficulty in explaining why
the size of such a correction 
 should coincide accurately with the
size of the QED correction which arises from much lower scales.
Up to date, 
no mechanism has been proposed to solve this
problem.

Among various existing models which attempt to explain origins
of Koide's mass formula, we find a class of models particularly
attractive \cite{Koide:1989jq}.
These are the models which predict the mass matrix of 
the charged leptons to be
proportional to the square of 
the vacuum expectation value (VEV)
of a 
scalar field (we denote it as $\Phi$) written in 
a 3--by--3 matrix form:
\bea
{\cal M}_\ell \propto \langle \Phi \rangle \langle \Phi \rangle
\, .
\label{MasPhisq}
\eea
Thus, $(\sqrt{m_e},\sqrt{m_\mu},\sqrt{m_\tau})$ is proportional
to the diagonal elements of $\langle \Phi \rangle$ in the
basis where it is diagonal.
The VEV $\langle \Phi \rangle$
is determined by minimizing the potential of scalar fields
in each model.
Hence, the origin of Koide's formula is
attributed to the specific form of the potential which realizes
this relation in the vacuum configuration.
Up to now, no model is complete with respect to symmetry:
Every model requires either
absence or strong suppression of
some of the terms in the potential (which are allowed by
the symmetry of that model),
without 
justification.

In this paper, we propose a possible
mechanism for cancelling the QED
correction to Koide's mass formula,
in the context of a theory with
family (horizontal) gauge symmetry.
In our study
we adopt the mechanism eq.~(\ref{MasPhisq}) for generating the charged
lepton masses at tree level, due to the following
reasons.
First, models with this mechanism are suited for
perturbative analyses.
Secondly, since $\Phi$ is renormalized multiplicatively,
the structure of radiative corrections is simple.

We consider an effective theory with family gauge symmetry
which
is valid up to some cut--off scale denoted by $\Lambda\,\,(\gg M_W)$.
Within such an effective theory, the charged lepton
masses may be induced by a higher--dimensional operator
\bea
{\cal O}=\frac{\kappa(\mu)}{\Lambda^2}\, 
\bar{\psi}_{Li}\, \Phi_{ik}\, \Phi_{kj}\, \varphi \, e_{Rj} \, .
\label{HigerDimOp1}
\eea
Here, $\psi_{Li}=(\nu_{Li},e_{Li})^T$ denotes the left--handed lepton 
$SU(2)_L$ doublet
of the $i$--th generation;
$e_{Rj}$ denotes the right-handed charged lepton
of the $j$--th generation;
$\varphi$ denotes the Higgs doublet field;
$\Phi$ is a 9--component scalar field
and is singlet under the Standard Model (SM) gauge group.
We suppressed all the indices except for the generation (family)
indices $i,j,k=1,2,3$.
(Summation over repeated indices is understood throughout
the paper.)
The dimensionless Wilson coefficient of this operator is
denoted as $\kappa(\mu)$.
Once $\Phi$ acquires a VEV, the operator $\cal O$ will
effectively be rendered to
the Yukawa interactions of the SM; 
after the Higgs field 
also acquires a VEV, $\langle\varphi\rangle=(0,v_{\rm ew}/\sqrt{2})^T$
with $v_{\rm ew}\approx 250$~GeV, the operator will induce the
charged--lepton mass matrix of the form
eq.~(\ref{MasPhisq}) at tree level:
\bea
{\cal M}_\ell^{\rm tree} = \frac{\kappa\,v_{\rm ew}}{\sqrt{2}\Lambda^2} 
\langle \Phi \rangle \langle \Phi \rangle
\, .
\eea

We consider radiative corrections
to the above mass matrix by the family gauge interaction.
First we consider the case, in which the gauge group is $SU(3)$
and both $\psi_L$ and $e_R$ are assigned to the $\bf 3$ (fundamental
representation) 
of this symmetry group.
With this choice of representation,  however, Koide's
formula is subject to a severe radiative correction unless the 
family gauge
interaction is strongly suppressed.
In fact, the 1--loop correction by the family gauge bosons induces an effective 
operator
\bea
{\cal O}' \sim \frac{\alpha_F}{\pi}\times{\kappa}\,
\bar{\psi}_{Li} \, \varphi \, e_{Ri} 
\times \frac{\langle\Phi\rangle_{jk}
\langle\Phi\rangle_{kj}
}{\Lambda^2} 
\, ,
\eea
hence corrections universal to all the charged--lepton masses,
$(\delta m_e, \delta m_\mu, \delta m_\tau) \propto (1,1,1)$,
are induced.
This is due to the fact that the dimension--4
operator $\bar{\psi}_{Li} \, \varphi \, e_{Ri} $
is not prohibited by symmetry.
Here, $\alpha_F=g_F^2/(4\pi)$ denotes the gauge
coupling constant
of the family gauge interaction.
As noted above, corrections which are proportional to individual
masses do not affect Koide's formula;
oppositely, the universal correction violates Koide's formula strongly.
In order that the correction to Koide's formula cancel the QED correction,
a naive estimate shows that 
$\alpha_F/\pi$ should be order $10^{-5}$, provided that
the cut--off $\Lambda$ is not too large and that the above operator
${\cal O}'$ is absent at tree level.
If ${\cal O}'$ exists at tree level, there should be a
fine tuning between the tree and 1--loop
contributions.
The situation is similar if the family symmetry is 
$O(3)$ and 
both $\psi_L$ and $e_R$ are in the $\bf 3$.
In these cases
\cite{Antusch:2007re} we were unable to find any sensible reasoning 
for the cancellation between the QED correction and
the correction induced by family gauge interaction, other than 
to regard the cancellation as just
a pure coincidence.
Hence, we will not investigate these choices of
representation further.

In the case that $\psi_L$ is assigned to $\bf 3$ and
$e_R$ to $\bar{\bf 3}$ (or {\it vice versa}) of $U(3)$
family gauge group,
(i) the dimension--4 operator
$\bar{\psi}_{Li} \, \varphi \, e_{Ri} $ is
prohibited by symmetry, and hence
corrections universal to all the three
masses do not appear; and
(ii) marked resemblance of the
radiative correction to the QED correction
follows.
We show these points explicitly in a specific setup.

We denote the generators for the fundamental representation
of $U(3)$ by $T^\alpha$
($0\leq\alpha\leq 8$), which satisfy
\bea
{\rm tr}\left(T^\alpha T^\beta\right)=\frac{1}{2}\, \delta^{\alpha\beta} 
~~~;~~~
T^\alpha = {T^\alpha}^\dagger
\, .
\label{U3generators}
\eea
$T^0=\frac{1}{\sqrt{6}}{\bf 1}$ is the generator of $U(1)$,  
while $T^a$ ($1\leq a \leq 8$) are the generators of
$SU(3)$.

We assign 
$\psi_L$ to the representation
$({\bf 3},1)$, where $\bf 3$ stands
for the $SU(3)$ representation and 1 for the $U(1)$ charge,
while $e_R$ is assigned to $(\bar{\bf 3},-1)$.
Under $U(3)$, the 9--component field
$\Phi$ transforms as three $({\bf 3},1)$'s.
Explicitly the transformations of these fields 
are given by
\bea
\psi_L \to U \, \psi_L \, ,
~~~
e_R \to U^* \, e_R \, ,
~~~
\Phi \to U \, \Phi 
\label{U3transf}
\eea
with $U = \exp \left(i\theta^\alpha T^\alpha\right)$.

We assume that the 
charged--lepton mass matrix is induced by a higher--dimensional
operator ${\cal O}^{(\ell)}$ similar to ${\cal O}$ in eq.~(\ref{HigerDimOp1}).
We further assume that $\langle \Phi \rangle$ can be brought to a
diagonal form 
in an appropriate basis.
Thus, in this basis ${\cal O}^{(\ell)}$, after
$\Phi$ and $\varphi$ acquire VEVs, turns to the lepton mass terms 
as
\bea
&&
{\cal O}^{(\ell)}\to 
\frac{\kappa^{(\ell)}(\mu)\,v_{\rm ew}}{\sqrt{2}\Lambda^2} \, 
 \bar{\psi}_{L}\,\Phi_d(\mu)^2\, \,
e_R \, ,
\label{DiagonalMassMat}
\eea
where
\bea
\Phi_d (\mu)=
\left(\begin{array}{ccc}
v_1(\mu)&0&0\\
0&v_2(\mu)&0\\
0&0&v_3(\mu)
\end{array}\right) \, ,
~~~ v_i(\mu) >0
\, .
\label{Phid}
\eea
When all $v_i$ are different,
$U(3)$ symmetry is completely broken by $\langle \Phi \rangle=\Phi_d$,
and the spectrum of the $U(3)$ gauge bosons is determined 
by $\Phi_d$.

Note that the operator ${\cal O}$ in eq.~(\ref{HigerDimOp1})
is {\it not} invariant under the $U(3)$ transformations
eq.~(\ref{U3transf}).
As an example of ${\cal O}^{(\ell)}$, one may consider 
\bea
{\cal O}^{(\ell)}_1=
\frac{\kappa^{(\ell)}(\mu)}{\Lambda^2}\,
\bar{\psi}_{L}\, \Phi\, \Phi^T\, \varphi \, e_{R} \, .
\label{exampleO1}
\eea
It is invariant under a larger symmetry  
$U(3)\times O(3)$, under which
$\Phi$ transforms as $\Phi\to U\Phi O^T$ ($O\,O^T = {\bf 1}$).
In this case, 
we need to assume, for instance, that the $O(3)$ symmetry is gauged 
and spontaneously
broken at a high energy
scale before the breakdown of the $U(3)$ symmetry, in order to
eliminate Nambu--Goldstone bosons and to
suppress mixing of the $U(3)$ and $O(3)$ gauge bosons.
In any case, 
the properties of ${\cal O}^{(\ell)}$ given by
eqs.~(\ref{DiagonalMassMat}) and (\ref{Phid})
are sufficient for computing the radiative correction
by the $U(3)$ gauge bosons to the mass matrix,
without an explicit form of ${\cal O}^{(\ell)}$.

We take the $U(1)$ and $SU(3)$ gauge coupling
constants to be the same:
\bea
\alpha_{U(1)}=\alpha_{SU(3)}=\alpha_F\, .
\label{UnivCouplings}
\eea
We compute the radiative correction in Landau gauge,
which is known to be convenient for computations in theories
with spontaneous symmetry breaking.
Through standard computation, one obtains
\bea
&&
\delta m^{\rm pole}_i =
-\frac{3\,\alpha_F}{8\,\pi}\left[
\log\left(
\frac{\mu^2}{v_i(\mu)^2}
\right) + c
\right] \,
{m}_i(\mu) \, ,
\label{alphaFcorr}
\\
&&
{m}_i(\mu) = 
\frac{\kappa^{(\ell)}(\mu)\,v_{\rm ew}}{\sqrt{2}\Lambda^2} \, v_i(\mu)^2 \, .
\label{mmu}
\eea
Here, $c$ is a constant independent of $i$.
The Wilson coefficient
$\kappa^{(\ell)}(\mu)$ is defined in $\overline{\rm MS}$ scheme.
$v_i(\mu)$ are defined as follows:
The VEV of $\Phi$ at renormlization
scale $\mu$, $\Phi_d(\mu)=\langle \Phi(\mu)\rangle$
given by eq.~(\ref{Phid}),
is determined by minimizing the 1--loop effective potential
in Landau gauge 
(although we do not discuss the explicit form of the effective
potential);
$\Phi$ is renormalized in $\overline{\rm MS}$ scheme.
We ignored terms suppressed by
${m}_i^2/v_j^2(\ll 1)$ in the above expression.
Note that the
pole mass is renormalization--group invariant and
gauge independent.
Therefore, the above expression is
rendered gauge independent if we
express $v_i(\mu)$ in terms of gauge independent parameters, 
such as coupling constants defined in on--shell scheme.

The form of the radiative correction given by eqs.~(\ref{alphaFcorr})
and (\ref{mmu})
is constrained by symmetries and
their breaking patterns.
As the diagonal elements of the VEV,
$v_3>v_2>v_1>0$, are successively turned on,
gauge symmetry is broken according to the 
pattern:
\bea
U(3) \to U(2)\to U(1)\to \mbox{nothing} \, .
\label{SymBreakPat}
\eea
At each stage, the gauge bosons corresponding to 
the broken generators
acquire masses and decouple.
Furthermore, the vacuum $\Phi_d$ and the 
family gauge interaction
respect a global 
$U(1)^3$
symmetry generated by
\bea
&&
\psi_L \to V\,\psi_L  ,
~
e_R \to V^*\, e_R ,
~
\Phi_d \to V\, \Phi_d V^* \, ,
\eea
with
\bea
V=
\left(\begin{array}{ccc}
e^{i\phi_1}&0&0\\
0&e^{i\phi_2}&0\\
0&0&e^{i\phi_3}
\end{array}\right) 
~~~;~~~
\phi_i \in {\bf R}
\, .
\eea
Although
${\cal O}^{(\ell)}$ after symmetry breakdown,
 eq.~(\ref{DiagonalMassMat}),
is not invariant under this transformation, the
variation can be absorbed into a redefinition of $v_i$.
As a result, the lepton mass matrix has a following
transformation property:
\bea
{\cal M}_\ell \biggr|_{v_i\to v_i \exp(i\phi_i)}
= V\, {\cal M}_\ell \, V^* \, .
\label{TransfU1Vcube}
\eea
This is satisfied including the 1--loop radiative correction.
The symmetry breaking pattern eq.~(\ref{SymBreakPat})
and the above transformation property 
constrain the form of the radiative correction
to $\delta m_i^{\rm pole} \propto v_i^2 [\log (|v_i|^2) + {\rm const.}]$,
where the constant is independent of $i$.
Note that $|v_i|^2$ in the argument of logarithm originates from
the gauge boson masses, which are invariant 
under $v_i \to v_i \exp(i\phi_i)$.

The universality of the $SU(3)$ and $U(1)$ gauge couplings 
eq.~(\ref{UnivCouplings}) is necessary to
guarantee the above symmetry breaking pattern eq.~(\ref{SymBreakPat}).
One may worry about validity of the assumption for the universality,
since the two couplings are renormalized differently in general.
The universality can be ensured approximately if these two 
symmetry groups are embedded
into a simple group down to a scale close to the relevant scale.
There are more than one ways to achieve this.
A simplest way would be to embed $SU(3)\times U(1)$
into $SU(4)$.
It is easy to verify that the $\bf 4$ of $SU(4)$ decomposes into
$({\bf 3}, -\frac{1}{2})\oplus ({\bf 1},\frac{3}{2})$ under $SU(3)\times U(1)$.
Hence, the $\bf 6$ (second-rank antisymmetric representation)
and $\bar{\bf 6}$ 
of $SU(4)$, respectively, include $(\bar{\bf 3},-1)$ and $({\bf 3}, 1)$.

Within the effective theory under consideration, the QED correction
to the pole mass is given 
just as in eq.~(\ref{QED1Lcorr}) with $\bar{m}_i(\mu)$
replaced by $m_i(\mu)$.
Recall that corrections of the form ${\rm const.}\times
{m}_i$ do not affect Koide's formula.
Noting $\log v_i^2 = \frac{1}{2}\log {m}_i^2 + \mbox{const.}$, 
one observes that if a relation between the QED and family gauge coupling constants
\bea
\alpha = \frac{1}{4}\, \alpha_F
\label{relalphas}
\eea
is satisfied,
the 1--loop radiative correction induced by family gauge interaction
cancels the 1--loop QED correction to Koide's mass formula.

Suppose the relation (\ref{relalphas}) is satisfied.
Then
\bea
m_i^{\rm pole} \propto v_i(\mu)^2
\label{RelPoleMassVEVs}
\eea
holds with a good accuracy.
This is valid for any value of $\mu$.
This means,
if $v_i(\mu)$ satisfy 
\bea
\frac{v_1(\mu)+v_2(\mu)+v_3(\mu)}
{\sqrt{v_1(\mu)^2+v_2(\mu)^2+v_3(\mu)^2}}
=\sqrt{\frac{3}{2}} 
\label{relvi}
\eea
at some scale $\mu$, Koide's formula is satisfied
at any scale $\mu$.
This is a consequence of the fact that $\Phi$
is multiplicatively renormalized.
Generally, the form of the effective potential varies with scale $\mu$.
If the relation (\ref{relvi}) is realized at some scale as a
consequence of a specific nature of the effective potential
(in Landau gauge), the same
relation holds
automatically at any scale.
Although these statements are formally true, 
physically one should consider scales only above the family
gauge boson masses, since decoupling of
the gauge bosons is not encoded 
in $\overline{\rm MS}$
scheme.
An interesting possibility is
to use eq.~(\ref{RelPoleMassVEVs}) to
relate the charged lepton pole masses with the VEV at the cut--off
scale, i.e.\ $\mu=\Lambda$, which sets a boundary (initial) 
condition of the effective theory.

The advantages of choosing Landau gauge in our computation are
two folds:
(1) The computation of the 1--loop
effective potential becomes particularly simple
(as well known in computations of the
effective potential in various models); in particular
there is no ${\cal O}(\alpha_F)$ correction to the effective
potential.
(2) The lepton wave--function renormalization is finite;
as a consequence, the $p_\mu \gamma^\mu$ part of the lepton self--energy
is independent of generation.
Due to the former property, there is no ${\cal O}(\alpha_F)$ correction to
the relation eq.~(\ref{relvi}) if it is satisfied at tree level.
Due to the latter property, 
the correction to Koide's formula is determined by
renormalization of ${\cal O}^{(\ell)}$ alone, and a simple relation
to $\langle\Phi(\mu)\rangle$ follows.

Let us comment on gauge dependence of our prediction.
If we take another gauge and express
the radiative correction $\delta m_i^{\rm pole}$
in terms of $\langle\Phi(\mu)\rangle$, 
the coefficient of $\log (\mu^2/\langle\Phi\rangle^2)$ changes, and
other non--trivial flavor dependent corrections 
are induced.
Suppose the relation eq.~(\ref{relvi}) is satisfied at tree
level \footnote{
To simplify the argument we consider only those gauges in which
tree--level vacuum configuration is gauge independent, such as the class
of gauges considered in \cite{DelCima:1999gg}.
}.
The VEV $\langle\Phi\rangle$ in another gauge
receives an ${\cal O}(\alpha_F)$ correction,
which induces a correction to eq.~(\ref{relvi}) at  ${\cal O}(\alpha_F)$.
These additional corrections to $\delta m_i^{\rm pole}$ at
${\cal O}(\alpha_F)$ should cancel altogether 
if they are reexpressed in terms of the tree--level $v_i$'s which
satisfy eq.~(\ref{relvi}),
since 
the ${\cal O}(\alpha_F)$ correction to the relation (\ref{relvi}) 
vanishes in Landau gauge.
General analyses on gauge dependence of 
the effective potential may be found in \cite{b}.

Now we speculate on a possible scenario how the relation
(\ref{relalphas}) may be satisfied.
The scale of $\alpha$ is determined by the charged
lepton masses, while the scale of $\alpha_F$ is determined by 
the family gauge boson masses, which should be much
higher than the electroweak 
scale.
Since the relevant scales of the two couplings
are very different, we are unable to 
avoid assuming some
accidental factor (or parameter tuning) to achieve this condition.
Instead we seek for an indirect evidence which indicates such an
accident has occurred in Nature.
The relation (\ref{relalphas}) shows that the value of
$\alpha_F$ is close to that of the
weak gauge coupling constant $\alpha_W$, since $\sin^2\theta_W(M_W)$ is close
to $1/4$.
In fact, within the SM, $\frac{1}{4}\,\alpha_W(\mu)$ approximates $\alpha(m_\tau)$
at scale $\mu \sim 10^2$--$10^3$~TeV.
Hence, if the electroweak $SU(2)_L$ gauge
group and the $U(3)$ family gauge
group are unified 
around this scale, naively we expect that
$
\alpha \approx \frac{1}{4}\, \alpha_F
$
is satisfied.
Since $\alpha_W$ runs relatively slowly in the SM, 
even if the unification scale is varied within a factor of 3,
Koide's mass formula is satisfied within the present experimental
accuracy.
This shows the level of parameter tuning required in this scenario.

We assume that anomalies introduced by the couplings
of fermions to family gauge bosons are cancelled, which
requires existence of
fermions other than the SM fermions.
Furthermore, we assume that
all the additional fermions acquire
masses of order $\langle \Phi \rangle$ or larger after the
spontaneous breakdown of $U(3)$, so that
they decouple from the SM sector at and below the electroweak scale.

A characteristic 
prediction of the present scenario is the
existence of lepton flavor violating processes at 
$10^2$--$10^3$~TeV scale.
For instance, assuming that the down--type
quarks are in the same representation
of $U(3)$ as the charged leptons, 
and that the mass matrices of the charged leptons and
down--type quarks
are simultaneously diagonalized in an appropriate basis,
we find
$
\Gamma(K_L\to e\mu)\approx
{m_\mu^2m_{K_L}f_K^2}
/({16\pi v_2^4})
$.
Comparing to the present experimental bound 
${\rm Br}(K_L\to e\mu)<4.7\times 10^{-12}$ \cite{Amsler:2008zz},
we obtain a limit $v_2\simgt 5\times 10^2$~TeV.
Naively this limit may be marginally in
conflict with the estimated
unification scale in the above scenario.
This depends, however, 
rather heavily on our assumptions on the quark
sector.
In the case that there exist additional 
factors in the quark sector which suppress the decay width
by a few orders of magnitude,
we may expect
 a signal
for $K_L\to e\mu$ not far beyond the present experimental
reach.
Predictions concerning purely leptonic processes
are less model dependent, but expected event rates are far below
present experimental sensitivities.

Models, which predict a realistic
charged lepton spectrum incorporating the 
mechanism proposed in this paper, will be discussed elsewhere
\cite{fff}.


The author is grateful to K.~Tobe for discussion.
This work is supported in part by Grant-in-Aid for
scientific research No.\ 17540228 from
MEXT, Japan.


\end{document}